\documentclass[preprint,pra,showpacs,nofootinbib]{revtex4}
\usepackage{mathrsfs}
\usepackage{float,epsfig}
\usepackage{graphicx}% Include figure files
\usepackage{dcolumn}% Align table columns on decimal point
\usepackage{bm}% bold math
\usepackage{amsmath,amssymb,amsthm}
\usepackage[colorlinks=true,linkcolor=blue]{hyperref}
\usepackage{subfigure}
\usepackage{booktabs}
\usepackage[mathscr]{euscript}

\newcommand{\bea}{\begin{eqnarray}}
\newcommand{\eea}{\end{eqnarray}}
\newcommand{\beq}{\begin{equation}}
\newcommand{\eeq}{\end{equation}}

\def\/{\over}

\newcommand{\ket}[1]{|#1\rangle}

\begin{document}
\title{Entanglement dynamics for uniformly accelerated two-level atoms coupled with electromagnetic vacuum fluctuations}
\author{  Yiquan Yang$^{1}$, Jiawei Hu$^{1}$\footnote{Corresponding author: hujiawei@nbu.edu.cn} and Hongwei Yu$^{
1,2}$\footnote{Corresponding author: hwyu@hunnu.edu.cn }}

\affiliation{
$^{1}$Center for Nonlinear Science and Department of Physics, Ningbo University,  Ningbo, Zhejiang 315211, China\\
$^2$Department of Physics and Synergetic Innovation Center for Quantum Effects and Applications, Hunan Normal University, Changsha, Hunan 410081, China}

\begin{abstract}
We investigate the entanglement dynamics of two uniformly accelerated atoms with the same acceleration perpendicular to their separation. The two-atom system is treated as an open system coupled with fluctuating electromagnetic fields in the Minkowski vacuum, and in the Born-Markov approximation the master equation that describes  the completely positive time evolution of the two-atom system is derived. In particular, we investigate the phenomena of entanglement degradation, generation, revival and enhancement. As opposed to the scalar-field case, the entanglement dynamics is crucially dependent on the polarization directions of the atoms. For the two-atom system with certain acceleration and separation, the polarization directions of the atoms may determine whether entanglement generation, revival or enhancement happens, while for entanglement degradation, they affect the decay rate of entanglement. A comparison between the entanglement evolution of accelerated atoms and that of static ones immersed in a thermal bath at the Unruh temperature shows that they are the same only when the acceleration is extremely small.

\end{abstract}


\maketitle

\section{Introduction}

Entanglement, one of the most intriguing features that distinguish the classical from the quantum worlds, was introduced by Schr\"{o}dinger in his discussion of the foundations of quantum mechanics \cite{sch}. Quantum entanglement is of interest because it is not only crucial to our understanding of quantum mechanics, but also lies at the heart of many novel applications, such as quantum information and computation  \cite{information1,information2}. However, decoherence that arises from the inevitable interaction with external environments may degrade the entanglement between quantum systems, and it is one of the main obstacles to the realization of quantum information technologies \cite{information3,information4}. Therefore, the entanglement dynamics of open quantum systems is an important issue in quantum information science. Recently, it has been found that two initially entangled atoms can become separable within a finite time much shorter than the local decoherence time, which is known as entanglement sudden death \cite{esd1,esd2}, and has been verified experimentally \cite{esd-3}. On the other hand, due to the indirect interactions from the common bath in which the atoms are immersed, entanglement can also be created from initially separable atoms~\cite{ent-bath1,ent-bath2,ent-birth,ent-bath3}, and the destroyed entanglement may be recreated~\cite{ent-rev}. For a two-atom system with a nonvanishing separation immersed in a thermal bath, entanglement sudden birth requires an appropriate interatomic separation and a sufficiently small bath temperature while entanglement sudden death is a general feature \cite{ent-bath3}. When the interatomic separation is vanishing, entanglement can persist in the asymptotic equilibrium state for atoms in certain initial states \cite{Benatti-job}.

A uniformly accelerated observer perceives the Minkowski vacuum as a thermal bath with a temperature proportional to its proper acceleration; this is known as the Unruh effect~\cite{Unruh}. Therefore, it is of interest to see how the entanglement between accelerating atoms behave. Benatti and Floreanini investigated the entanglement generation for two uniformly accelerated atoms which are infinitely close to each other weakly interacting with a bath of fluctuating scalar fields in the Minkowski vacuum and found that the asymptotic equilibrium state turns out to be the same as that in a thermal bath at the Unruh temperature, thus shedding light on the Unruh effect from a new perspective \cite{Benatti-pra}. This work was generalized to the case of two accelerating atoms with a finite interatomic separation near a reflecting boundary, and it has been shown that the conditions for entanglement generation at the beginning of evolution are not exactly the same as those in a thermal bath \cite{yu-prd}. The whole evolution process of entanglement for accelerating atoms has also been examined in Refs. \cite{Matsas,Doukas1,BLHu,Hu-pra}.

The studies in Refs.~\cite{Benatti-pra,yu-prd,Matsas,Doukas1,Hu-pra,BLHu} model the environment the atoms coupled to as a bath of fluctuating scalar fields. A more realistic model would be  a bath of fluctuating vacuum electromagnetic fields. For such a model, it has been shown that the spontaneous emission rate \cite{Takagi,em2,em3,em4} and the Lamb shift \cite{lamb1,lamb2} of an accelerated atom are not completely equivalent to those in a thermal bath, in contrast to the scalar-field case. In Ref. \cite{Doukas2}, the entanglement evolution of two maximally entangled electron spins has been investigated, with one undergoing uniform acceleration in the presence of electromagnetic fields, and the other at rest and isolated from fields.

In the present paper, we plan to study the entanglement dynamics of two mutually independent two-level atoms accelerating with the same acceleration perpendicular to the separation coupled with the fluctuating electromagnetic fields in the Minkowski vacuum, and compare the results with those of static atoms immersed in a thermal bath at the Unruh temperature. In addition to entanglement degradation and generation, which were also examined in our previous work in which the environment is modeled as a bath of fluctuating scalar fields \cite{Hu-pra}, we also consider entanglement revival and enhancement for atoms prepared in different initial states. In particular, we investigate how the entanglement dynamics is influenced by the polarization of the atoms, which do not play any role in the scalar-field case.

%%%%%%%%%%%%%%%%%%%%%%%%%%

\section{The Master Equation}

We consider a two-atom system weakly coupled to a bath of fluctuating electromagnetic fields in the Minkowski vacuum. The Hamiltonian of the whole system can be expressed as follows
\begin{equation}
 H=H_A+H_F+H_I\;.
\end{equation}
Here $H_A$ is the Hamiltonian of the two-atom system,
\begin{equation}
H_A={\omega\over 2}\,\sigma_3^{(1)}+
    {\omega\over 2}\,\sigma_3^{(2)}\;,
\end{equation}
where $\sigma_i^{(1)}=\sigma_i\otimes{\sigma_0}$, $\sigma_i^{(2)}={\sigma_0}\otimes\sigma_i$, with $\sigma_i~(i=1,2,3)$ being the Pauli matrices and $\sigma_0$ the $2\times2$ unit matrix, and $\omega$ is the energy level spacing of the atoms. $H_F$ denotes the Hamiltonian of the electromagnetic fields and its explicit expression is not required here. $H_I$ describes the interaction between the two-atom system and the fluctuating electromagnetic fields, which can be written in the multipolar coupling scheme as
\beq\label{HI}
H_I(\tau)=-\textbf{D}^{(1)}(\tau)\cdot\textbf{E}(x^{(1)}(\tau))
          -\textbf{D}^{(2)}(\tau)\cdot\textbf{E}(x^{(2)}(\tau))\;,
\eeq
where $\textbf{D}^{(\alpha)}(\tau)~(\alpha=1,2)$ is the electric-dipole moment, and ${\bf E}(x^{(\alpha)}(\tau))$ denotes the electric-field strength. Here we have ignored the dipole-dipole interaction between the two atoms.

Under the Born-Markov approximation, the quantum master equation which governs the evolution of the reduced density matrix of the two-atom system can be derived. In the  derivation, we define two Lindblad operators
\begin{equation}
\textbf{A}^{(\alpha)}(\omega)\equiv \textbf{A}^{(\alpha)}=\textbf{d}^{(\alpha)}\sigma_{-},\,\textbf{A}^{(\alpha)}(-\omega)\equiv \textbf{A}^{(\alpha)\dag}=\textbf{d}^{(\alpha)*}\sigma_{+},
\end{equation}
where $\textbf{d}^{(\alpha)}=\langle 0|\textbf{D}^{(\alpha)}|1\rangle$ is the transition matrix element of the dipole operator of the $\alpha$th atom. In the interaction picture the atomic dipole operator can thus be written as
\begin{equation}
\textbf{D}^{(\alpha)}(\tau)=\textbf{d}^{(\alpha)}\sigma_{-} e^{-i{\omega}{\tau}}+\textbf{d}^{(\alpha)*}\sigma_{+} e^{i{\omega}{\tau}}.
\end{equation}
If we suppose that the two-atom system and the fluctuating electromagnetic fields are uncorrelated at the beginning, the initial state can then be written as $\rho_{\text{tot}}(0)=\rho(0)\otimes\rho_{\text{f}}(0)$,
where $\rho_{\text{f}}(0)$ denotes the Minkowski vacuum state of the fields, and $\rho(0)$ denotes the initial state of the two-atom system. In the weak-coupling limit, the master equation of the two-atom system can be obtained in the Kossakowski-Lindblad form as \cite{Lindblad,Lindblad2,open}
\begin{equation}\label{master}
{\partial\rho(\tau)\/\partial\tau}
=-i\big[H_{\rm eff},\,\rho(\tau)\big]+{\cal L}[\rho(\tau)]\,,
\end{equation}
where
\begin{equation}
H_{\rm eff}
=H_A-\frac{i}{2}\sum_{\alpha,\beta=1}^2\sum_{i,j=1}^3
H_{ij}^{(\alpha\beta)}\,\sigma_i^{(\alpha)}\,\sigma_j^{(\beta)}\,,
\end{equation}
and
\begin{equation}
{\cal L}[\rho]
={1\over2} \sum_{\alpha,\beta=1}^2\sum_{i,j=1}^3
 C_{ij}^{(\alpha\beta)}
 \big[2\,\sigma_j^{(\beta)}\rho\,\sigma_i^{(\alpha)}
 -\sigma_i^{(\alpha)}\sigma_j^{(\beta)}\, \rho
 -\rho\,\sigma_i^{(\alpha)}\sigma_j^{(\beta)}\big]\,.
\end{equation}
Here $C_{ij}^{(\alpha\beta)}$ and
$H_{ij}^{(\alpha\beta)}$ are determined by the Fourier and Hilbert transforms, ${\cal G}^{(\alpha\beta)}_{mn}(\lambda)$ and  ${\cal K}^{(\alpha\beta)}_{mn}(\lambda)$, of the electromagnetic field correlation functions
\begin{equation}\label{green}
G^{(\alpha\beta)}_{mn}(\tau-\tau')
=\langle E_{m}(\tau,\mathbf{x}_{\alpha}) E_{n}(\tau',\mathbf{x}_\beta)
 \rangle\;,
\end{equation}
 which are defined as
\begin{equation}\label{fourierG}
{\cal G}^{(\alpha\beta)}_{mn}(\lambda)
=\int_{-\infty}^{\infty} d\Delta\tau\,
 e^{i{\lambda}\Delta\tau}\, G^{(\alpha\beta)}_{mn}(\Delta\tau)\; ,
\end{equation}
\begin{equation}
{\cal K}^{(\alpha\beta)}_{mn}(\lambda)
=\frac{P}{\pi i}\int_{-\infty}^{\infty} d\omega\
 \frac{{\cal G}^{(\alpha\beta)}_{mn}(\omega)}{\omega-\lambda} \;,
\end{equation}
with $P$ denoting the principal value. Then $C_{ij}^{(\alpha\beta)}$ can be written explicitly as
\beq\label{C}
C_{ij}^{(\alpha\beta)}
= A^{(\alpha\beta)}\delta_{ij}
 -iB^{(\alpha\beta)}\epsilon_{ijk}\,\delta_{3k}
 -A^{(\alpha\beta)}\delta_{3i}\,\delta_{3j}\;,
\eeq
where
\begin{equation}\label{AB}
\begin{aligned}
A^{(\alpha\beta)}
={1\/4}\,[\,{\cal G}^{(\alpha\beta)}(\omega)
 +{\cal G}^{(\alpha\beta)}(-\omega)]\;,\\
B^{(\alpha\beta)}
={1\/4}\,[\,{\cal G}^{(\alpha\beta)}(\omega)
 -{\cal G}^{(\alpha\beta)}(-\omega)]\;,
\end{aligned}
\end{equation}
and
\begin{equation}\label{G}
{\cal G}^{(\alpha\beta)}(\omega)
=\sum_{m,n=1}^3\, d^{(\alpha)}_m d^{(\beta)*}_n\,
{\cal G}^{(\alpha\beta)}_{mn}(\omega)\;.
\end{equation}
Similarly, $H^{(\alpha\beta)}_{ij}$ can be obtained by replacing ${\cal
G}^{(\alpha\beta)}_{mn}$ with ${\cal
K}^{(\alpha\beta)}_{mn}$ in the above equations..

\section{entanglement dynamics of the two-atom system}

In this section we investigate the entanglement evolution of the two-atom system accelerating with the same acceleration perpendicular to the separation and compare it with that of static ones immersed in a thermal bath at the Unruh temperature.

The trajectories of the two uniformly accelerated atoms are
\beq
\begin{aligned}\label{traj}
t_1(\tau)&={1\/a}\sinh a\tau\;, &
x_1(\tau)&={1\/a}\cosh a\tau\;, &
y_1(\tau)&=0\;, &
z_1(\tau)&=L\;,\\
t_2(\tau)&={1\/a}\sinh a\tau\;, &
x_2(\tau)&={1\/a}\cosh a\tau\;, &
y_2(\tau)&=0\;, &
z_2(\tau)&=0\;,
\end{aligned}
\eeq
respectively. The correlation function of electromagnetic fields in the Minkowski vacuum takes the form
\begin{eqnarray}\label{correlation}
\langle0|E_m(x(\tau))E_m(x(\tau'))|0\rangle&=&{1\/4\pi^2}\;(\partial
_0\partial_0^\prime\delta_{mn}-\partial_m\partial_n^\prime)\nonumber\\&&\times
{1\/(x-x')^2+(y-y')^2+(z-z')^2-(t-t'-i\varepsilon)^2}\;.
\end{eqnarray}
Now we calculate the correlation functions of accelerating two-atom system from the general formula (\ref{correlation}) with a Lorentz transformation from the laboratory frame to the frame of the atoms, which can be expressed as
\beq
G^{(11)}_{mn}(x,x')=G^{(22)}_{mn}(x,x')
=\frac{a^4}{16\pi^2}\frac{1}{\sinh^4(\frac{au}{2}-i\epsilon)}\delta_{mn}\;,
\eeq
and
\beq
\begin{aligned}
G^{(\alpha\beta)}_{mn}(x,x')
=&\frac{a^4}{16\pi^2}\frac{1}{\left[\sinh^2(\frac{au}{2}-i\epsilon)-{a^2L^2\/4}\right]^3}
 \bigg\{[\delta_{mn}+aL{\varepsilon_{\alpha\beta3}}(\textit{l}_{m}\textit{k}_{n}-\textit{l}_{n}\textit{k}_{m})]{\sinh^2 \frac{au}{2}}\\
&+\frac{a^2L^2}{4}\bigg[(\delta_{mn}-2\textit{l}_{m}\textit{l}_{n}){\cosh^2 \frac{au}{2}}+(\delta_{mn}-2\textit{l}_{m}\textit{l}_{n}-2\textit{k}_{m}\textit{k}_{n}){\sinh^2 \frac{au}{2}}\bigg]\bigg\},
\end{aligned}
\eeq
for $\alpha \neq \beta$, where $u=\tau-\tau'$, $k_\mu=(0, 1, 0, 0)$ and $l_\mu=(0, 0, 0, 1)$.

The Fourier transforms of the above correlation functions are
\beq
{\cal G}^{(11)}_{mn}(\lambda)=G^{(22)}_{mn}(\lambda)
=\frac{1}{3\pi}{\lambda}^3 \left(1+\frac{1}{e^{2\pi\lambda/a}-1}\right)
f^{(11)}(\lambda,a) \delta_{mn} \;,
\eeq
where
\begin{equation}
\label{fs1}
f^{(11)}(\lambda,a)
=1+\frac{a^2}{\lambda^2}\;,
\end{equation}
and
\beq
{\cal G}^{(\alpha\beta)}_{mn}(\lambda)
=\frac{1}{3\pi}{\lambda}^3 \left(1+\frac{1}{e^{2\pi\lambda/a}-1}\right) f^{(\alpha\beta)}_{mn}(\lambda,a,L)\;,
\eeq
for $\alpha \neq \beta$, where
\begin{eqnarray}\label{f11}
&&\begin{aligned}
f^{(12)}_{11}(\lambda,a,L)
=&f^{(21)}_{11}(\lambda,a,L)=\frac{12}{{{\lambda ^3}{L^3}{{(4 + {a^2}{L^2})}^{5/2}}}}\\
&\times\bigg\{2\lambda L(1 + {a^2}{L^2}){(4 + {a^2}{L^2})^{1/2}}\cos \left(\frac{{2\lambda }}{a}{\sinh ^{ - 1}}\frac{{aL}}{2}\right)\\
&-[4-4{\lambda ^2}{L^2} + {a^2}{L^2}(2-{\lambda ^2}{L^2}+{a^2}{L^2})]\sin \left(\frac{{2\lambda }}{a}{\sinh ^{-1}}\frac{{aL}}{2}\right)\bigg\}\;,
\end{aligned}\;\\
&&\begin{aligned}
f^{(12)}_{22}(\lambda,a,L)
=&f^{(21)}_{22}(\lambda,a,L)=\frac{3}{{{\lambda ^3}{L^3}{{(4 + {a^2}{L^2})}^{3/2}}}}\\
&\times\bigg[\lambda L(2 + {a^2}{L^2}){(4 + {a^2}{L^2})^{1/2}}\cos \left(\frac{{2\lambda }}{a}{\sinh ^{ - 1}}\frac{{aL}}{2}\right)\\
&+(-4+4{\lambda ^2}{L^2} + {a^2}{\lambda ^2}{L^4})\sin \left(\frac{{2\lambda }}{a}{\sinh ^{-1}}\frac{{aL}}{2}\right)\bigg]\;,
\end{aligned}\;\\
&&\begin{aligned}
f^{(12)}_{33}(\lambda,a,L)
=&f^{(21)}_{33}(\lambda,a,L)=-\frac{3}{{{\lambda ^3}{L^3}{{(4 + {a^2}{L^2})}^{5/2}}}}\\
&\times\bigg\{\lambda L(16 + 2{a^2}{L^2}+{a^4}{L^4}){(4 + {a^2}{L^2})^{1/2}}\cos \left(\frac{{2\lambda }}{a}{\sinh ^{ - 1}}\frac{{aL}}{2}\right)\\
&+[-32+{a^4}{\lambda ^2}{L^6} + 4{a^2}{L^2}(-5+{\lambda ^2}{L^2})]\sin \left(\frac{{2\lambda }}{a}{\sinh ^{-1}}\frac{{aL}}{2}\right)\bigg\}\;,
\end{aligned}\;\\
&&\begin{aligned}\label{f13}
f^{(12)}_{13}(\lambda,a,L)
=&-f^{(12)}_{31}(\lambda,a,L)=-f^{(21)}_{13}(\lambda,a,L)=f^{(21)}_{31}(\lambda,a,L)=-\frac{6a}{{{\lambda ^3}{L^2}{{(4 + {a^2}{L^2})}^{5/2}}}}\\
&\times\bigg\{\lambda L(-2 + {a^2}{L^2}){(4 + {a^2}{L^2})^{1/2}}\cos \left(\frac{{2\lambda }}{a}{\sinh ^{ - 1}}\frac{{aL}}{2}\right)\\
&+[4+4{\lambda ^2}{L^2} + {a^2}{L^2}(4+{\lambda ^2}{L^2})]\sin \left(\frac{{2\lambda }}{a}{\sinh ^{-1}}\frac{{aL}}{2}\right)\bigg\}\;,
\end{aligned}
\end{eqnarray}
with other components being zero.

In this paper, we assume that the magnitudes of the electric dipoles of the atoms are the same, i.e.,
$|\mathbf{d}^{(1)}|=|\mathbf{d}^{(2)}|=|\mathbf{d}|$, but the orientations may be different. The coefficients of the dissipator in the master equation (\ref{master}) can then be calculated according to Eqs. (\ref{C})-(\ref{G})
\begin{eqnarray}
&&C_{ij}^{(11)}=C_{ij}^{(22)}
=A_1\,\delta_{ij}-iB_1\epsilon_{ijk}\,\delta_{3k}
 -A_1\delta_{3i}\,\delta_{3j}\;,\\
&&C_{ij}^{(12)}=C_{ij}^{(21)}
=A_2\,\delta_{ij}-iB_2\epsilon_{ijk}\,\delta_{3k}
 -A_2\delta_{3i}\,\delta_{3j}\;,
\end{eqnarray}
where
\begin{eqnarray}\label{abc}
&&A_1=\frac{\Gamma_0}{4} f^{(11)}(\omega,a)\coth{\pi\omega\/a}\;,\\
&&A_2=\frac{\Gamma_0}{4} \sum_{i,j=1}^3 f^{(12)}_{ij}(\omega,a,L)\hat{d}^{(1)}_i\hat{d}^{(2)}_j\coth{\pi\omega\/a}\;,\label{a2}\\
&&B_1=\frac{\Gamma_0}{4} f^{(11)}(\omega,a)\;,\\
&&B_2=\frac{\Gamma_0}{4} \sum_{i,j=1}^3 f^{(12)}_{ij}(\omega,a,L)\hat{d}^{(1)}_i\hat{d}^{(2)}_j\;,\label{b2}
\end{eqnarray}
with $\Gamma_0=\; \omega^3 |\mathbf{d}|^2 /3\pi$ being the spontaneous emission rate, and $\hat{d}^{(\alpha)}_{i}$ being a unit vector defined as
$\hat{d}^{(\alpha)}_{i}=d^{(\alpha)}_{i}/|\mathbf{d}|$.

To describe the evolution of the two-qubit system, we choose to work in the coupled basis $\{|G\rangle=|00\rangle,|A\rangle={1\/\sqrt{2}}(|10\rangle-|01\rangle), |S\rangle={1\/\sqrt{2}}(|10\rangle+|01\rangle),|E\rangle=|11\rangle\}$ for convenience. Then a set of equations which are decoupled from other matrix elements can be derived as \cite{ent-states}
\bea\label{evolution}
&\dot{\rho}_{GG}=-4(A_1-B_1)\rho_{GG}+2(A_1+B_1-A_2-B_2)\rho_{AA}+2(A_1+B_1+A_2+B_2)\rho_{SS}\;,\label{rho-gg}\\
&\dot{\rho}_{AA}=-4(A_1-A_2)\rho_{AA}+2(A_1-B_1-A_2+B_2)\rho_{GG}+2(A_1+B_1-A_2-B_2)\rho_{EE}\;,\label{rho-aa}\\
&\dot{\rho}_{SS}=-4(A_1+A_2)\rho_{SS}+2(A_1-B_1+A_2-B_2)\rho_{GG}+2(A_1+B_1+A_2+B_2)\rho_{EE}\;,\label{rho-ss}\\
&\dot{\rho}_{EE}=-4(A_1+B_1)\rho_{EE}+2(A_1-B_1-A_2+B_2)\rho_{AA}+2(A_1-B_1+A_2-B_2)\rho_{SS}\;,\label{rho-ee}\\
&\dot{\rho}_{AS}=-4 A_1 \rho_{AS}\;,\dot{\rho}_{SA}=-4 A_1 \rho_{SA}\;,\label{rho-as}\\
&\dot{\rho}_{GE}=-4 A_1 \rho_{GE}\;,\dot{\rho}_{EG}=-4 A_1 \rho_{EG}\;,\label{rho-ge}
\eea
where $\rho_{IJ}=\langle I|\rho|J \rangle$. Therefore, if we assume that the initial density matrix takes the X form, i.e., the nonzero elements are arranged along the main diagonal and antidiagonal of the density matrix in the decoupled basis $\{|00\rangle,|01\rangle,|10\rangle,|11\rangle\}$, the X structure will be preserved during the evolution.

We take the concurrence introduced by Wootters \cite{concurrence1} as a measurement of quantum entanglement, which ranges from 0 (for separable states) to 1 (for maximally entangled states). For the X states, the concurrence takes the form \cite{concurrence2}
\beq\label{concurrence}
C[\rho(\tau)]=\max\{0,K_1(\tau),K_2(\tau)\}\;,
\eeq
where
\bea\label{k1}
&&K_1(\tau)=\sqrt{[\rho_{AA}(\tau)-\rho_{SS}(\tau)]^2
-[\rho_{AS}(\tau)-\rho_{SA}(\tau)]^2}-2\sqrt{\rho_{GG}(\tau)\rho_{EE}(\tau)}\;,\\
&&K_2(\tau)=2 |\rho_{GE}(\tau)|- \sqrt{[\rho_{AA}(\tau)+\rho_{SS}(\tau)]^2-[\rho_{AS}(\tau)+\rho_{SA}(\tau)]^2}\;.\label{k2}
\eea
With the help of this formula, it is obvious that, for atoms with a nonvanishing separation, the asymptotic state is separable, since
\beq
\begin{aligned}
&\rho_{AA}(\infty)=\rho_{SS}(\infty)=
- \frac{{ - A_1^3 + {A_1}A_2^2 + {A_1}B_1^2 - {A_1}B_2^2}}{{4\left( {A_1^3 - {A_1}A_2^2 - {A_2}{B_1}{B_2} + {A_1}B_2^2} \right)}}\;,\\
&\rho_{AS}(\infty)=\rho_{SA}(\infty)=\rho_{GE}(\infty)=\rho_{EG}(\infty)= 0\;,
\end{aligned}
\eeq
and therefore $K_1(\infty)$ (\ref{k1}) and $K_2(\infty)$ (\ref{k2}) are negative. That is, entanglement sudden death is a general feature for accelerated atoms with a finite separation. When the separation approaches zero, the model is no longer valid because the dipole-dipole interaction would play an important part and the atoms cannot be even considered as distinguishable \cite{BLHu12}. Therefore, we do not discuss the case of vanishing separation in detail in this paper. We also note that, at very small separations, there are some discrepancies between atoms modeled as two-level systems \cite{BLHu12} and harmonic oscillators \cite{BLHu09}, which deserve further investigation.

Before a thorough comparison between the entanglement dynamics of accelerated atoms and static ones in a thermal bath, let us note that, for small acceleration, the entanglement dynamics of the uniformly accelerated atoms is essentially the same as that of the thermal case. To show this, we expand the functions $f^{(11)}(\lambda,a)$, $f^{(12)}_{11}(\lambda,a,L)$, $f^{(12)}_{22}(\lambda,a,L)$, $f^{(12)}_{33}(\lambda,a,L)$, and $f^{(12)}_{13}(\lambda,a,L)$ with respect to acceleration $a$ as
\begin{eqnarray}\label{fexpand1}
&&f^{(11)}(\lambda,a)
=1+O{\left[ a \right]^2}\;,\\
&&f^{(12)}_{11}(\lambda,a,L)\label{fexpand3}
=\frac{{3\lambda L\cos \lambda L - 3\sin \lambda L + 3{\lambda ^2}{L^2}\sin \lambda L}}{{2{\lambda ^3}{L^3}}}+ O{\left[ a \right]^2}\;,\\
&&f^{(12)}_{22}(\lambda,a,L)
=\frac{{3\lambda L\cos \lambda L - 3\sin \lambda L + 3{\lambda ^2}{L^2}\sin \lambda L}}{{2{\lambda ^3}{L^3}}}+ O{\left[ a \right]^2}\;,\\
&&f^{(12)}_{33}(\lambda,a,L)
=\frac{{ - 3\lambda L\cos \lambda L + 3\sin \lambda L}}{{{\lambda ^3}{L^3}}}+O{\left[ a \right]^2}\;,\\
&&f^{(12)}_{13}(\lambda,a,L)
=0+O{\left[ a \right]}\;.\label{fexpand2}
\end{eqnarray}
In the limit $a\to 0$, the functions (\ref{fexpand1})-(\ref{fexpand2}) are exactly the same as those in the thermal case, which can be calculated with the method of imaginary time \cite{QFTCS}. As the acceleration increases, the entanglement dynamics for uniformly accelerated atoms can generally be distinguished from that of the static ones in a thermal bath at the Unruh temperature. In the following we will address this issue in details.

\subsection{ Entanglement degradation}

First, we discuss the entanglement degradation of two-atom systems initially prepared in the symmetric state $\ket{S}$ and the antisymmetric state $\ket{A}$, both of which are maximally entangled.

When the interatomic separation is very large ($L\to\infty$), the modulating functions (\ref{f11})-(\ref{f13}), and thus $A_2$ and $B_2$, tend to zero. Therefore, the evolution of the populations ${\rho}_{AA}$ (\ref{rho-aa}) and ${\rho}_{SS}$ (\ref{rho-ss}) are the same, and there is no difference in the entanglement dynamics whether the initial state is $\ket{A}$ or $\ket{S}$, which agrees with the scalar-field case \cite{Hu-pra}. However, in the electromagnetic case, the entanglement dynamics for uniformly accelerated atoms can be distinguished from those immersed in a thermal bath at the corresponding Unruh temperature in the large separation limit in the sense that the decay rate of concurrence of accelerated atoms at $\tau=0$ is $\Gamma_0(1+a^2 /\omega^2)\coth{\pi\omega\/2a}$, while it is $\Gamma_0\coth{\pi\omega\/2a}$ for the thermal case, which is different from the scalar-field case~\cite{Hu-pra}. Note that the discrepancy between the dynamics of accelerated atoms and that of static ones in a thermal bath is not unique to the two-atom case, and it has already been shown in the study of the transition rate \cite{Takagi,em2,em3,em4} and the Lamb shift \cite{lamb1} of a single atom.

\begin{figure}[htbp]
\centering
\subfigure{\includegraphics[width=0.49\textwidth]{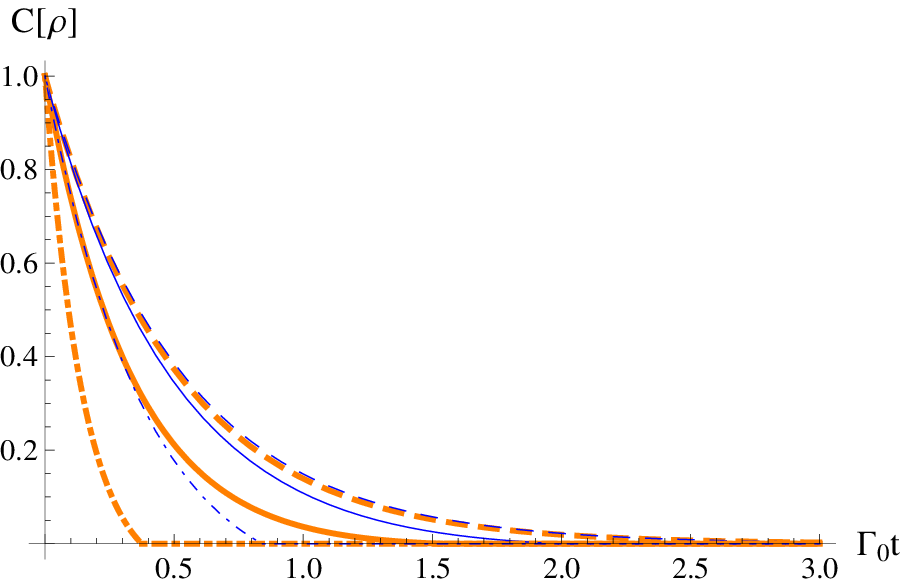}}
\subfigure{\includegraphics[width=0.49\textwidth]{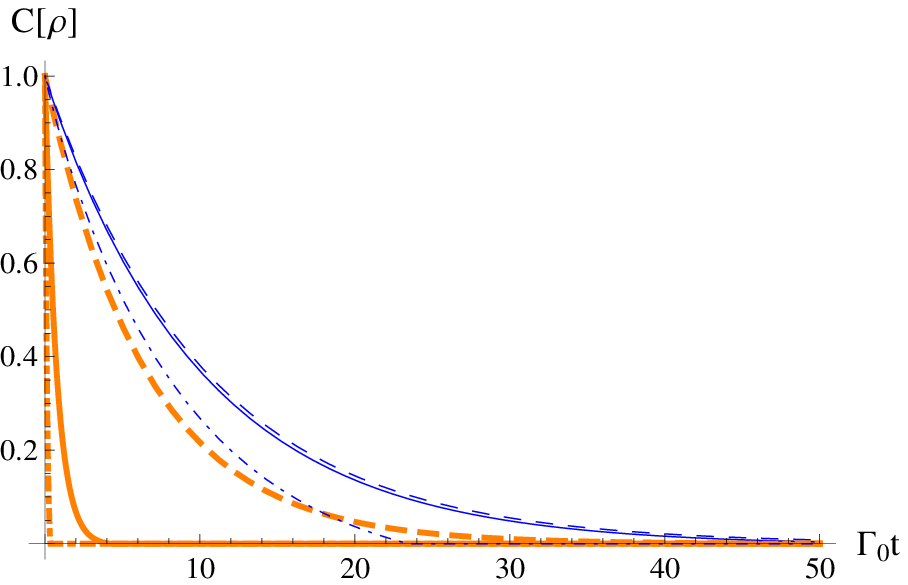}}
\caption{(Color online) Comparison between the dynamics of concurrence for uniformly accelerated atoms (bold orange  lines) and static atoms in a thermal bath at the Unruh temperature (fine blue lines) initially prepared in $|S\rangle$ (left) and $|A\rangle$ (right), with $\omega L=1$. Both of the two atoms are polarized along the positive $z$ axis. The dashed, solid, and dot-dashed lines correspond to $a/\omega=1/4$, $a/\omega=1$, and $a/\omega=2$, respectively.}
\label{pic-1}
\end{figure}

\begin{figure}[htbp]
\centering
\subfigure{\includegraphics[width=0.49\textwidth]{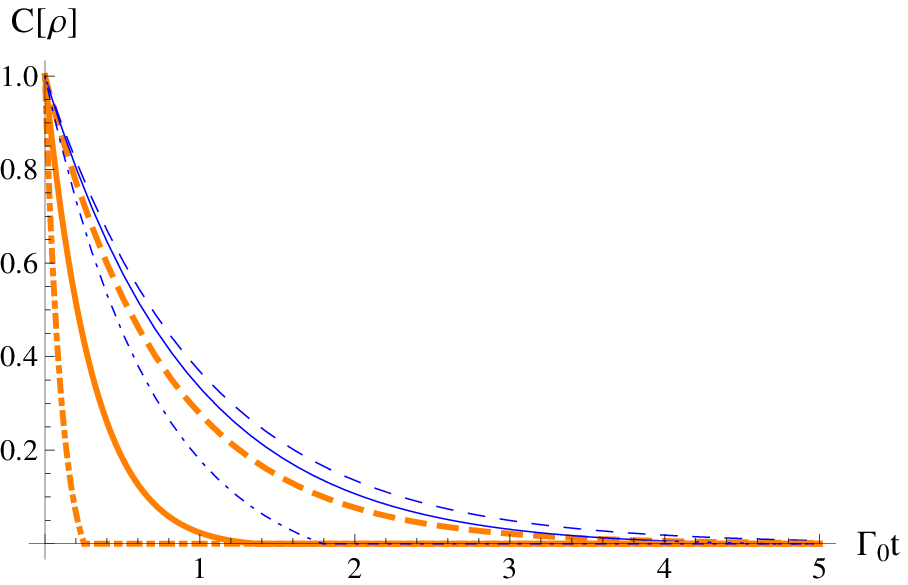}}
\subfigure{\includegraphics[width=0.49\textwidth]{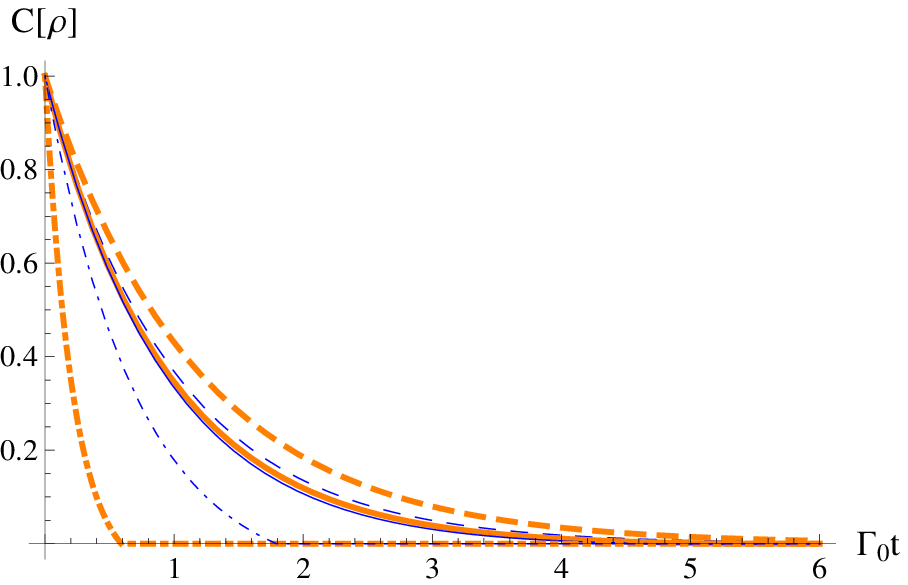}}
\caption{(Color online) Comparison between the dynamics of concurrence for uniformly accelerated atoms (bold orange lines) and static atoms in a thermal bath at the Unruh temperature (fine blue lines) initially prepared in  $|S\rangle$ (left) and $|A\rangle$ (right), with $\omega L=1$. The two atoms are polarized along the positive $z$ axis and the positive $x$ axis, respectively. The dashed, solid, and dot-dashed lines correspond to $a/\omega=1/4$, $a/\omega=1$, and $a/\omega=2$, respectively.}
\label{pic-2}
\end{figure}

For intermediate separations which are comparable with the transition wavelength of the atoms $(L\sim\omega^{-1})$, we numerically solve Eqs. (\ref{rho-gg})-(\ref{rho-ee}) as the solutions are rather complicated. In Figs. \ref{pic-1} and \ref{pic-2} we make a comparison between the dynamics of concurrence for uniformly accelerated atoms and static ones initially prepared in $\ket{A}$ and $\ket{S}$. First, we assume that both of the two atoms are polarized along the positive $z$ axis. Figure~\ref{pic-1} shows that the decay of the concurrence of accelerated atoms is always faster than that of the static ones in a thermal bath, no matter if the initial state is $\ket{A}$ or $\ket{S}$. Then we rotate the polarization of the atom located at $z=0$ towards the positive $x$ axis and leave the atom at $z=L$ unchanged, i.e., $\hat{\mathbf{d}}^{(1)}=(0,0,1)$ and $\hat{\mathbf{d}}^{(2)}=(1,0,0)$. In this case, for initial state $\ket{S}$, the concurrence of accelerated atoms decays faster than that of the static ones in a thermal bath, while if the two-atom system is initially in the state $\ket{A}$, the concurrence decays slower than that of the static ones, as shown in Fig. \ref{pic-2}. Note that if the polarization directions of the two atoms are exchanged, i.e., $\hat{\mathbf{d}}^{(1)}=(1,0,0)$ and $\hat{\mathbf{d}}^{(2)}=(0,0,1)$, the entanglement dynamics will be different, because in this case the contribution of the coefficients $A_2$ (\ref{a2}) and $B_2$ (\ref{b2}) comes from $f^{(12)}_{13}$, which is different from $f^{(12)}_{31}$ in the former case.

\subsection{ Entanglement generation}

Now we study the entanglement dynamics for two-atom systems initially prepared in a separable state $|E\rangle$. From Eqs. (\ref{concurrence})-(\ref{k2}) we know that entanglement generation happens only if the difference of populations between the symmetric and antisymmetric states $|\rho_{AA}-\rho_{SS}|$ outweighs the factor $2\sqrt{\rho_G\rho_E}$. Therefore, the entanglement generation of two-atom systems initially in the state $|E\rangle$ may happen after a finite time of evolution via spontaneous emission, which is known as the delayed sudden birth of entanglement \cite{ent-birth}.

\begin{figure}[htbp]
\centering
\subfigure{\includegraphics[width=0.49\textwidth]{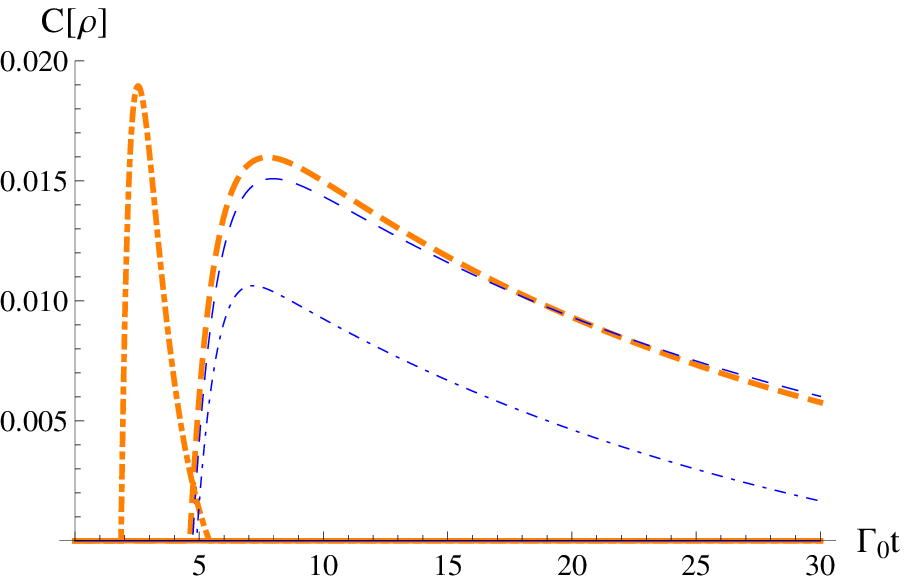}}
\subfigure{\includegraphics[width=0.49\textwidth]{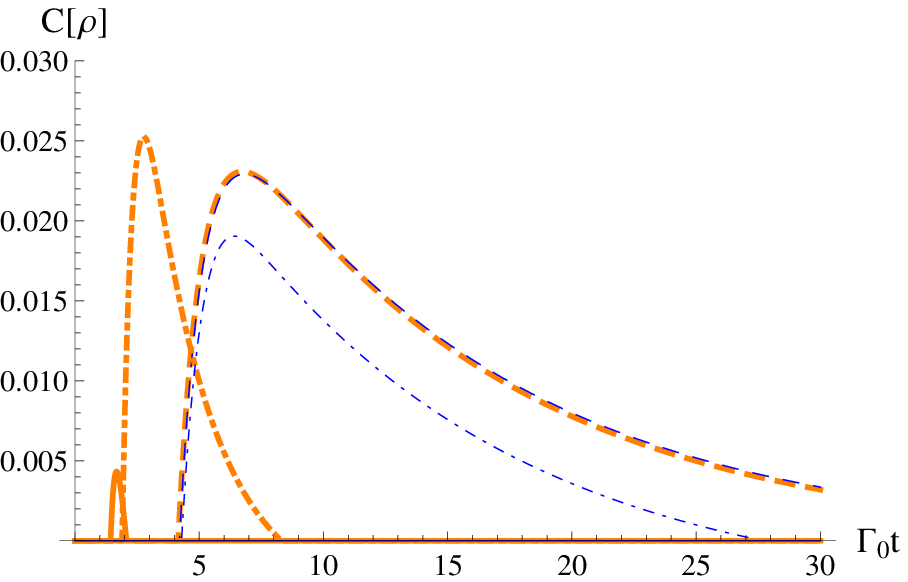}}
\caption{(Color online) Comparison between the dynamics of concurrence for uniformly accelerated atoms (thick orange lines) and static atoms in a thermal bath at the Unruh temperature (fine blue lines) initially prepared in $|E\rangle$, with $\omega L=2/3$. Both atoms are polarized along the positive $z$ axis (left) or the positive $y$ axis (right). The dashed, dot-dashed, and solid lines correspond to $a/\omega=1/10$, $a/\omega=1$, and $a/\omega=7/5$, respectively.}
\label{pic-3}
\end{figure}

From Fig. \ref{pic-3} we observe that the lifetime of entanglement decreases as acceleration increases. When the acceleration increases to $a/\omega=7/5$, entanglement generation does not happen for $z$-axis polarized atoms, while the $y$-axis polarized atoms can still be entangled. However, neither $z$-axis polarized nor $y$-axis polarized static atoms can get entangled if they were immersed in a thermal bath at the corresponding Unruh temperature.

\begin{figure}[htbp]
\centering
\subfigure{\includegraphics[width=0.49\textwidth]{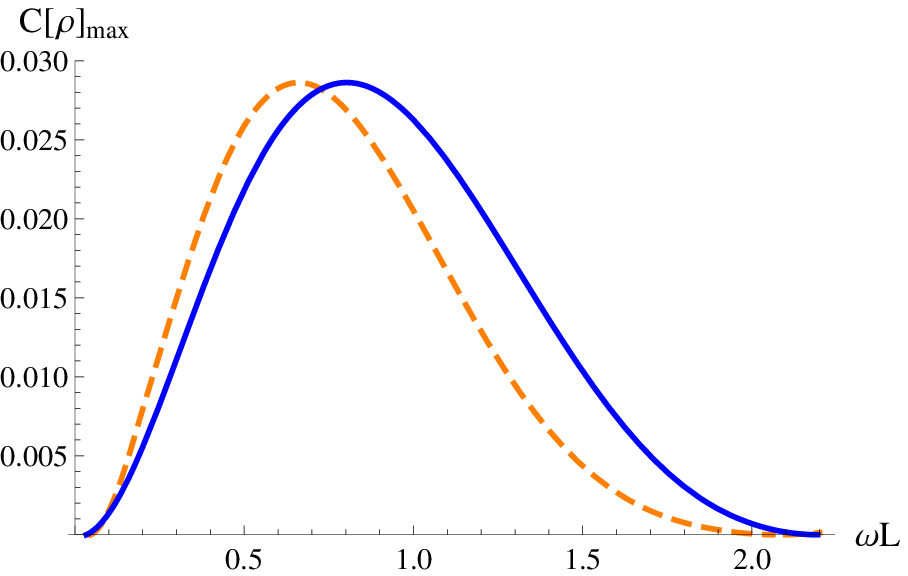}}
\subfigure{\includegraphics[width=0.49\textwidth]{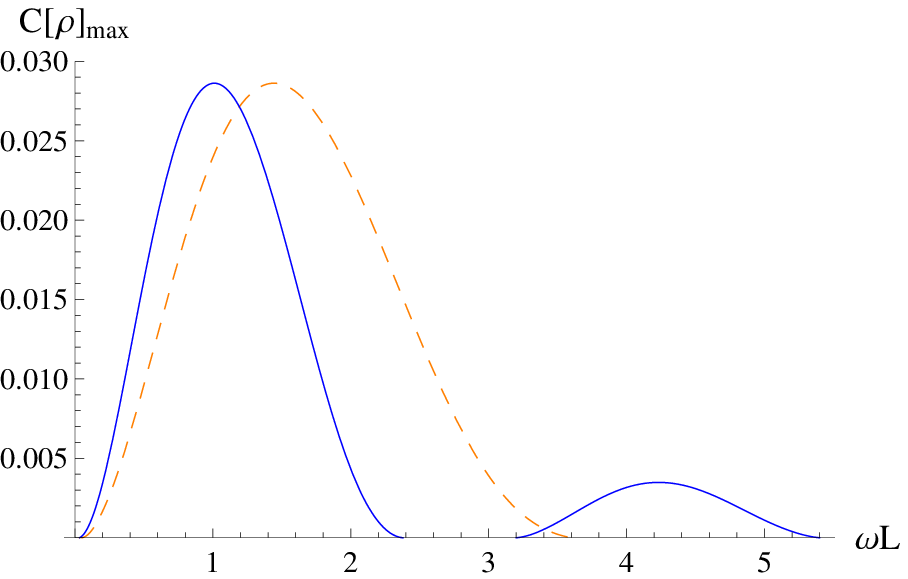}}
\caption{(Color online) Comparison between the maximum of concurrence during evolution for uniformly accelerated atoms (left) and static atoms in a thermal bath at the Unruh temperature (right) initially prepared in $|E\rangle$ with $a/\omega=2/3$. Both atoms are polarized along the positive $z$ axis (dashed orange lines) or the positive $y$ axis (solid blue lines). }
\label{pic-4}
\end{figure}
\begin{figure}[htbp]
\centering
\subfigure{\includegraphics[width=0.49\textwidth]{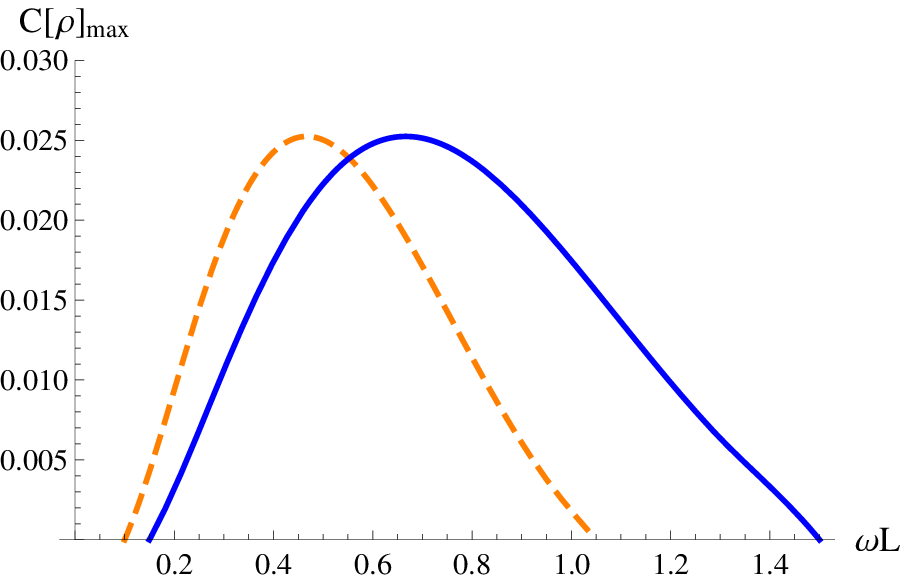}}
\subfigure{\includegraphics[width=0.49\textwidth]{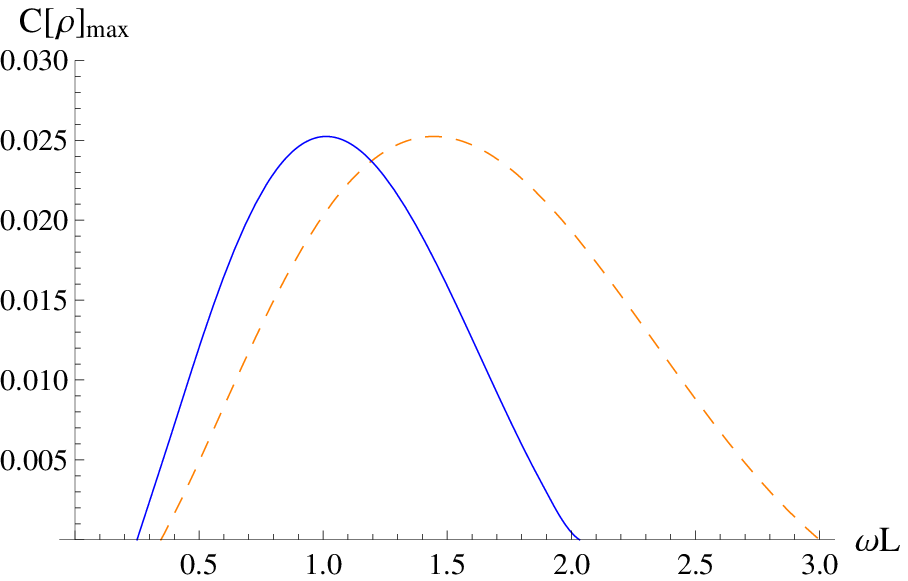}}
\caption{(Color online) Comparison between the maximum of concurrence during evolution for uniformly accelerated atoms (left) and static atoms in a thermal bath at the Unruh temperature (right) initially prepared in $|E\rangle$ with $a/\omega=1$. Both atoms are polarized along the positive $z$ axis (dashed orange lines) or the positive $y$ axis (solid blue lines). }
\label{pic-5}
\end{figure}

In Figs. \ref{pic-4} and \ref{pic-5}, we study the effects of atomic separation on the maximum of entanglement generated during the evolution. It is shown that there always exist a minimum (nonzero) and a maximum interatomic separation within which the atoms can be entangled for both the accelerated atoms and static ones in a thermal bath. Apart from acceleration (temperature), this interval is also dependent on the atomic polarization. For $y$-axis polarized static atoms immersed in a thermal bath with $a/\omega=2/3$, there is a dark interval that entanglement can not be created, as depicted in Fig. \ref{pic-4}, while there is no such dark interval for the corresponding accelerated case. In the case $a/\omega=1$, the intervals that permit entanglement generation for accelerated atoms are smaller than those in the thermal case.

\begin{figure}[htbp]
\centering
\subfigure{\includegraphics[width=0.49\textwidth]{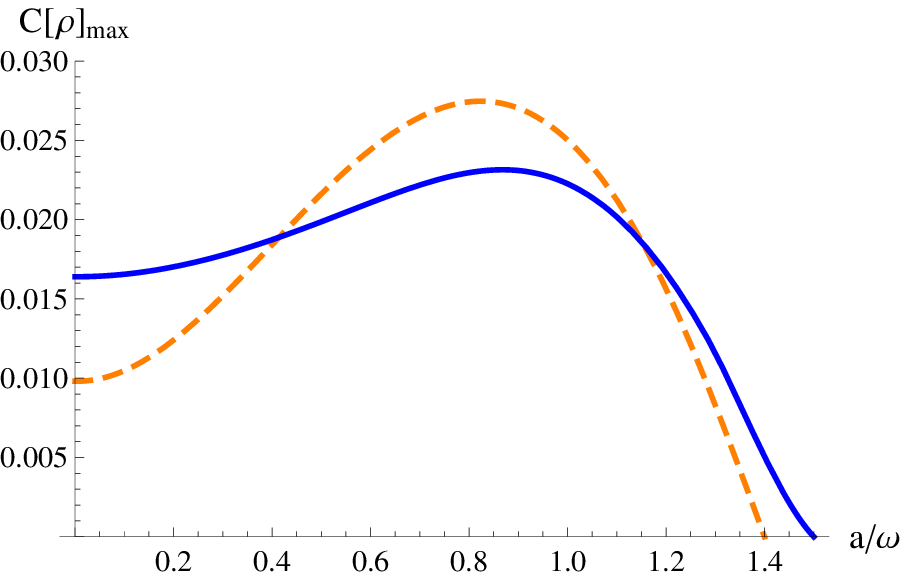}}
\subfigure{\includegraphics[width=0.49\textwidth]{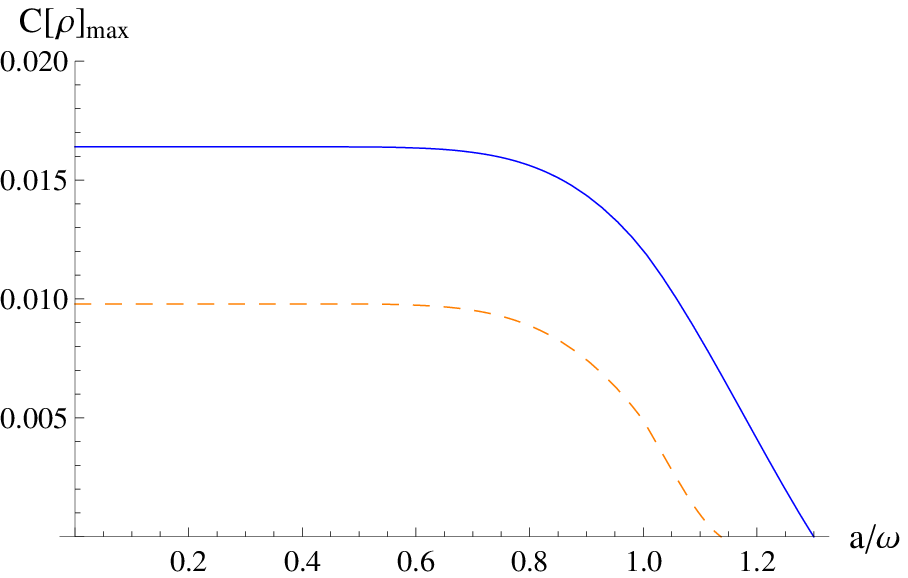}}
\caption{(Color online) Comparison between the maximum of concurrence during evolution for uniformly accelerated atoms (left) and static atoms in a thermal bath at the Unruh temperature (right) initially prepared in $|E\rangle$ with $\omega L=1/2$. Both atoms are polarized along the positive $z$ axis (dashed orange lines) or the positive $y$ axis (solid blue lines).}
\label{pic-6}
\end{figure}

Next we investigate the relation between the maximum concurrence during evolution and acceleration (temperature). As shown in Fig. \ref{pic-6}, for static atoms immersed in a thermal bath, the maximum of concurrence always decreases as the temperature increases, and the rate of change is extremely slow when the temperature is small, while for accelerated atoms, it may not be a monotonic function of acceleration, which is consistent with the scalar case~\cite{Hu-pra}. Also, both the maximum of concurrence for a given acceleration $a$ and separation $L$, and the maximum of acceleration $a$ larger than which entanglement can not happen for a given separation $L$ can be affected by the polarization direction, as shown in Fig. \ref{pic-6}.

%%%%%%%%%%%%%%%%%%%%%%%%%%%%%%%%%%%%

\subsection{ Entanglement revival and enhancement}

In this part, we investigate the phenomena of entanglement revival and enhancement for atoms in the following initial states
\begin{eqnarray}\label{state}
\begin{aligned}
&\ket{\psi_1}=
\sqrt{p}\;\ket{A}+\sqrt{1-p}\;\ket{S}\;\;(0<p<1\;,p\neq1/2),\\
&\ket{\psi_2}=
\sqrt{p}\;\ket{G}+\sqrt{1-p}\;\ket{E}\;\;(0<p<1),\\
\end{aligned}
\end{eqnarray}
both of which are entangled states.

\begin{figure}[htbp]
\centering
\subfigure{\includegraphics[width=0.49\textwidth]{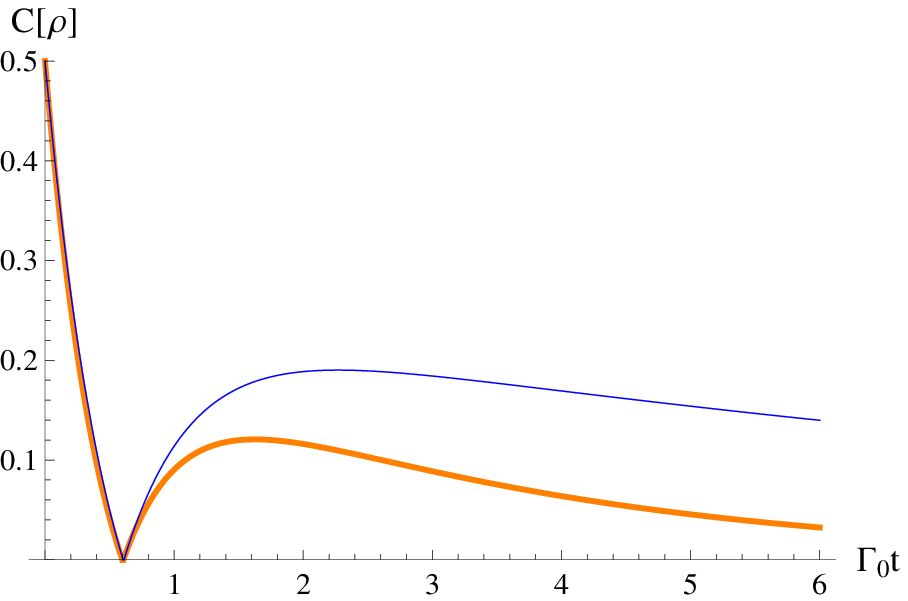}}
\subfigure{\includegraphics[width=0.49\textwidth]{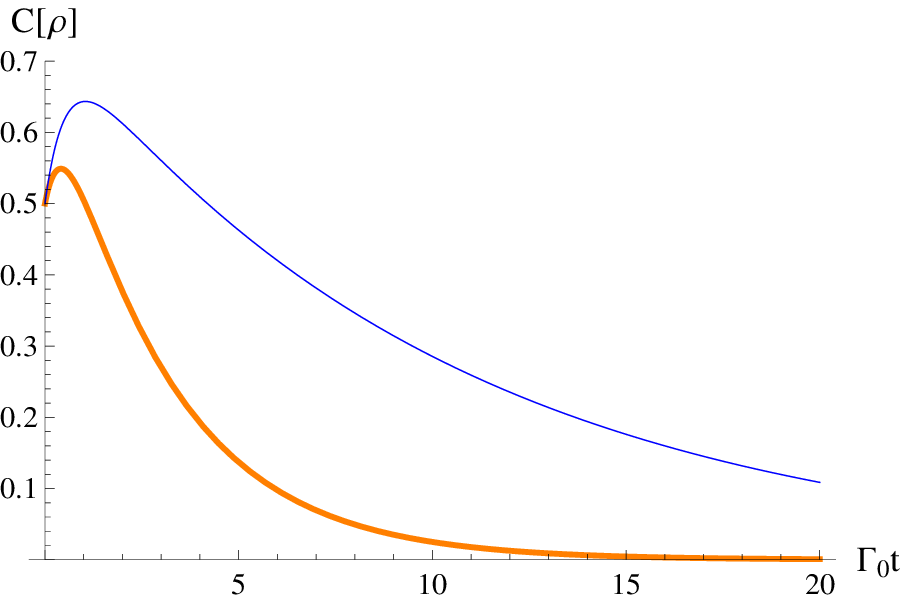}}
\caption{(Color online) Comparison between the dynamics of concurrence for uniformly accelerated atoms (thick orange  lines) and static atoms in a thermal bath at the Unruh temperature (fine blue lines) initially prepared in ${1\/2}\;\ket{A}+{\sqrt{3}\/2}\;\ket{S}$ (left) and ${\sqrt{3}\/2}\;\ket{A}+{1\/2}\;\ket{S}$ (right), with $a/\omega=1/2$ and $\omega L=1$. Both atoms are polarized along the positive $z$ axis.}
\label{pic-7}
\end{figure}

\begin{figure}[htbp]
\centering
\subfigure{\includegraphics[width=0.49\textwidth]{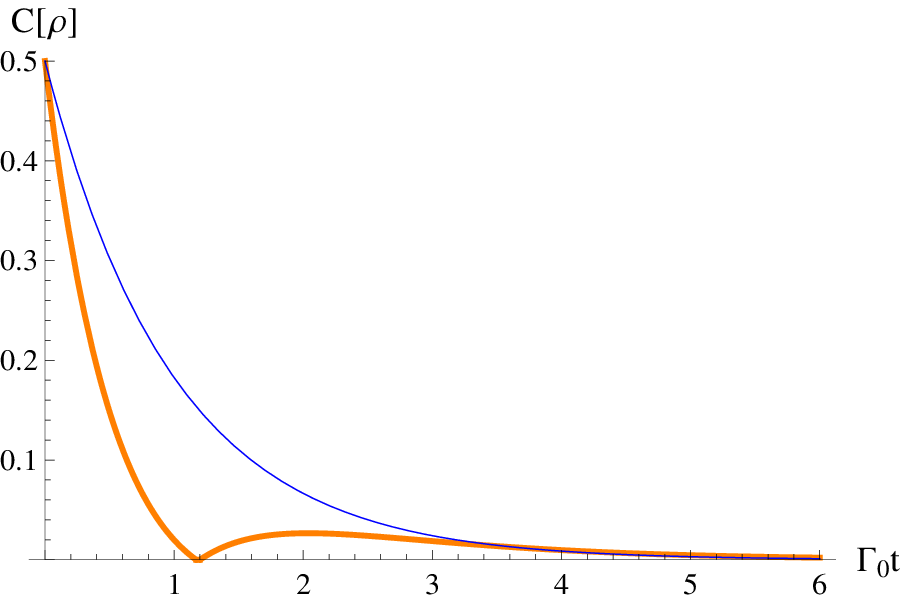}}
\subfigure{\includegraphics[width=0.49\textwidth]{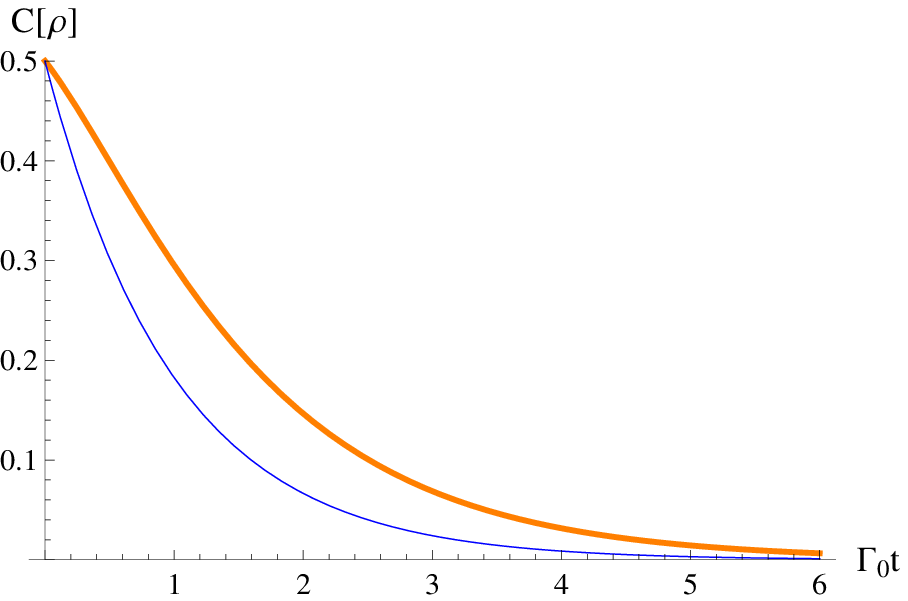}}
\caption{(Color online) Comparison between the dynamics of concurrence for uniformly accelerated atoms (thick orange lines) and static atoms in a thermal bath at the Unruh temperature (fine blue lines) initially prepared in ${1\/2}\;\ket{A}+{\sqrt{3}\/2}\;\ket{S}$ (left) and ${\sqrt{3}\/2}\;\ket{A}+{1\/2}\;\ket{S}$ (right), with $a/\omega=1/2$ and $\omega L=1$. The two atoms are polarized along the positive $z$ axis and the positive $x$ axis, respectively.}
\label{pic-8}
\end{figure}

\begin{figure}[htbp]
\centering
\subfigure{\includegraphics[width=0.49\textwidth]{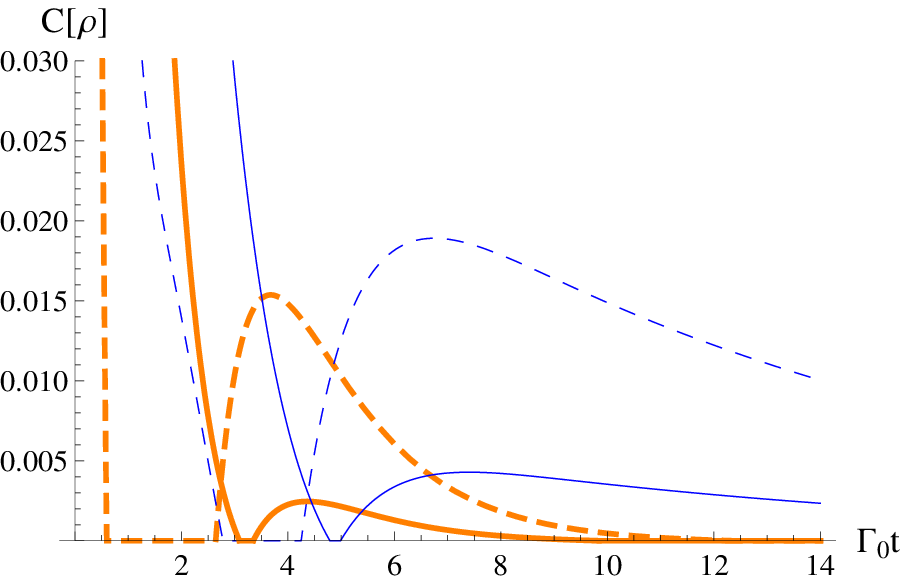}}
\subfigure{\includegraphics[width=0.49\textwidth]{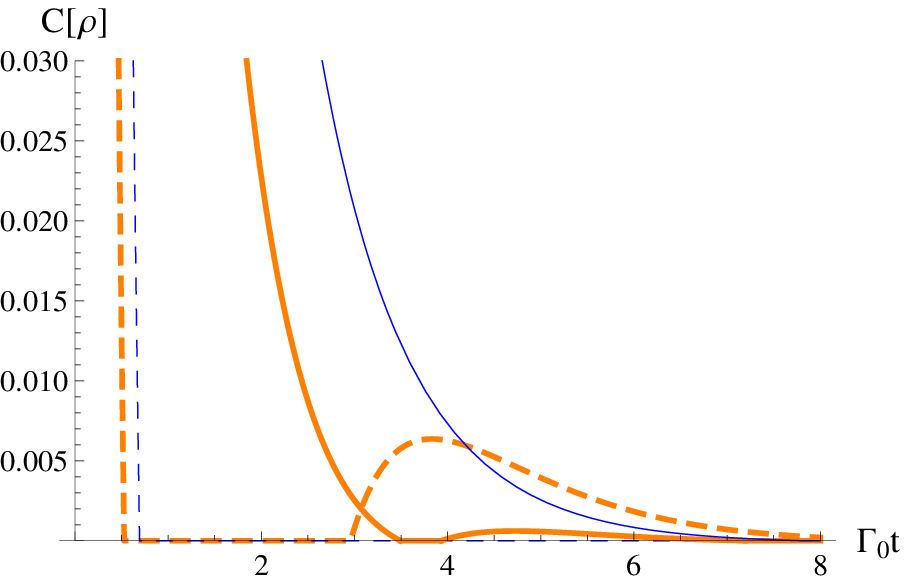}}
\caption{(Color online) Comparison between the dynamics of concurrence for uniformly accelerated atoms (thick orange lines) and static atoms in a thermal bath at the Unruh temperature (fine blue lines) initially prepared in $\ket{\psi_2}$, with $a/\omega=2/3$ and $\omega L=1$. Both atoms are polarized along the positive $z$ axis for the left part, while the two atoms are polarized along the positive $z$ axis and the positive $x$ axis, respectively, for the right part. The dashed and solid lines correspond to $p=1/5$ and $p=4/5$, respectively.}
\label{pic-9}
\end{figure}

In Figs. \ref{pic-7} and \ref{pic-8} we plot the time evolution of concurrence for atoms initially in $\ket{\psi_1}$, which is a superposition of $\ket{A}$ and $\ket{S}$. When both atoms are polarized along the positive $z$ axis, there are two different phenomena which are entanglement revival for $p=1/4$ and entanglement enhancement for $p=3/4$ for both the accelerated atoms and static ones in a thermal bath as shown in Fig.~\ref{pic-7}. However, when the polarization directions of the atoms are different such that atom 1 is polarized along the positive $z$ axis and atom 2 along the positive $x$ axis, for accelerated atoms entanglement revival happens when $p=1/4$ but the initial entanglement can not be enhanced when $p=3/4$, while for the thermal case there is neither entanglement revival nor enhancement, as shown in Fig. \ref{pic-8}. For atoms initially prepared in a superposition of $\ket{G}$ and $\ket{E}$, the destroyed entanglement can be revived for both the accelerated atoms and static ones in a thermal bath if both atoms are polarized along the positive $z$ axis, while if the two atoms are polarized along $z$ axis and $x$ axis, respectively, entanglement revival happens only for accelerated atoms, as shown in Fig. \ref{pic-9}. This is in accordance with the recent results obtained in Ref.~\cite{martin}, in which it has been shown that when the polarization directions are perpendicular, the atoms cannot harvest entanglement from the field.

\begin{figure}
\centering
\includegraphics[scale=1]{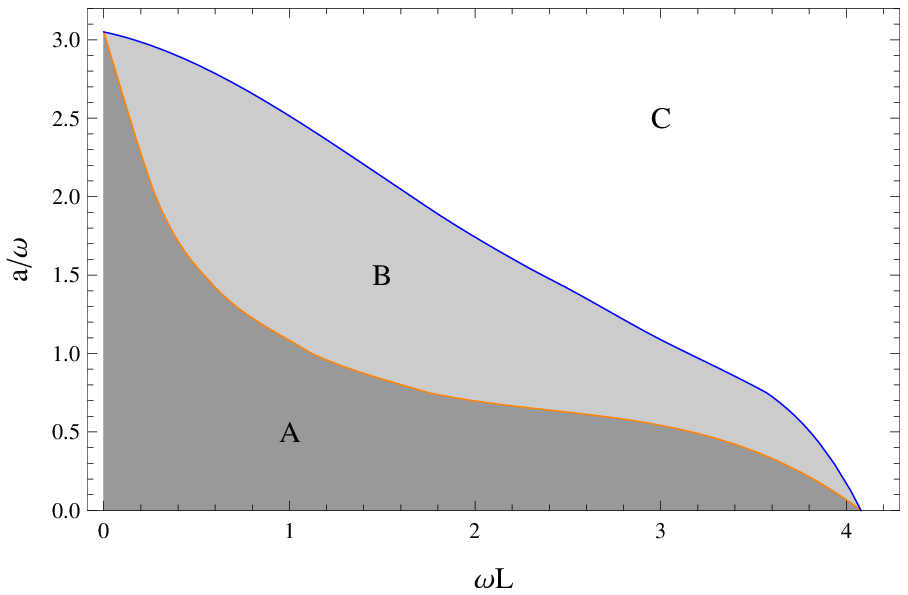}
\caption{(Color online) Entanglement profile for two-atom systems initially prepared in ${\frac{1}{2}}\ket{A}+{\frac{\sqrt3}{2}}\ket{S}$, both of which are polarized along the positive $z$ axis. Region A: Entanglement revival for both accelerated atoms and static atoms in a thermal bath. Region B: Entanglement revival for static atoms in a thermal bath. Region C: No entanglement revival. }
\label{pic-10}
\end{figure}
\begin{figure}
\centering
\includegraphics[scale=1]{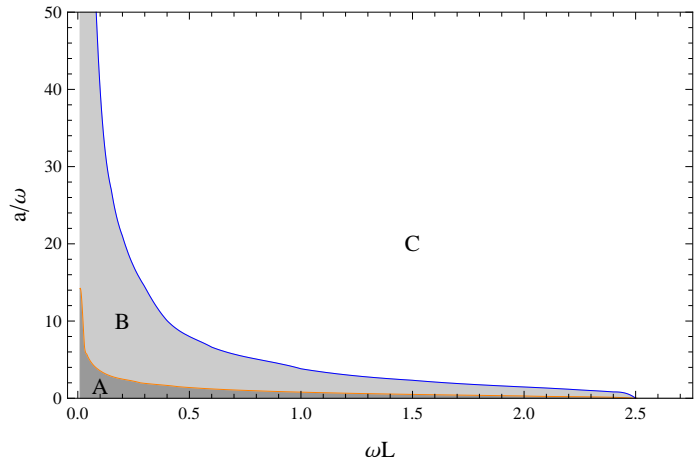}
\caption{(Color online) Entanglement profile for two-atom systems initially prepared in ${\frac{\sqrt3}{2}}\ket{A}+{\frac{1}{2}}\ket{S}$, both of which are polarized along the positive $z$ axis. Region A: Entanglement enhancement for both accelerated atoms and static atoms in a thermal bath. Region B: Entanglement enhancement for static atoms in a thermal bath. Region C: No entanglement enhancement. }
\label{pic-11}
\end{figure}
\begin{figure}
\centering
\includegraphics[scale=1]{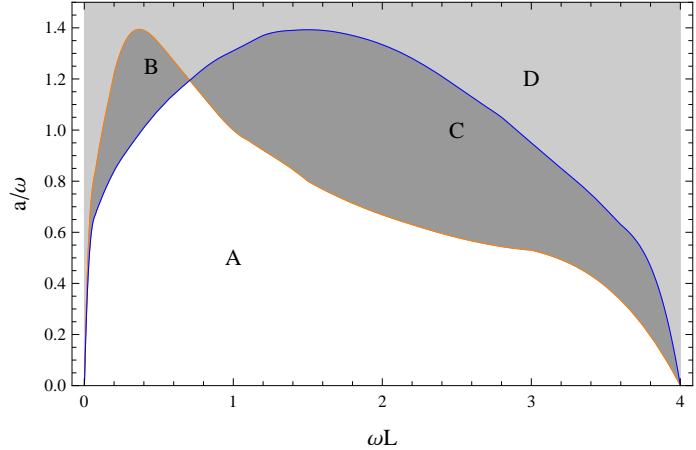}
\caption{(Color online) Entanglement profile for two-atom systems initially prepared in ${\frac{1}{\sqrt5}}\ket{G}+{\frac{2}{\sqrt5}}\ket{E}$, both of which are polarized along the positive $z$ axis. Region A: Entanglement revival for both accelerated atoms and static atoms in a thermal bath. Region B: Entanglement revival for accelerated atoms. Region C: Entanglement revival for static atoms in a thermal bath. Region D: No entanglement revival. }
\label{pic-12}
\end{figure}

In the following, we investigate the conditions for entanglement revival and enhancement for atoms initially prepared in
${\frac{1}{2}}\ket{A}+{\frac{\sqrt3}{2}}\ket{S}$, ${\frac{\sqrt3}{2}}\ket{A}+{\frac{1}{2}}\ket{S}$, and ${\frac{1}{\sqrt5}}\ket{G}+{\frac{2}{\sqrt5}}\ket{E}$, respectively. Here we assume that both atoms are polarized along the positive $z$ axis, while the conclusions are essentially the same if the polarizations of the two atoms are the same but towards a different direction. In Fig. \ref{pic-10}, it can be found that, for each separation, there exists an upper bound of acceleration larger than which entanglement revival does not happen. The region of entanglement revival for accelerated atoms is smaller than that of the static atoms in a thermal bath. However, they share the same upper bound of acceleration when interatomic separation tends to zero, and also the same upper bound of separation when acceleration tends to zero. Similarly, the conditions for  entanglement enhancement are as shown in Fig.~\ref{pic-11}. In contrast to the entanglement revival case (Fig. \ref{pic-10}), the upper bound of acceleration is much smaller than the corresponding Unruh temperature when the interatomic separation tends to zero. For atoms initially prepared in a superposition state of $\ket{G}$ and $\ket{E}$, the situation is quite different as shown in Fig. \ref{pic-12}. When the interatomic separation tends to zero, there is no entanglement revival for both the accelerated and static atoms in a thermal bath, and the possible region of entanglement revival for accelerated atoms is not a subset of that of the static ones, in contrast to the case when the initial state is a superposition state of $\ket{A}$ and $\ket{S}$.

\section{Conclusion}

In conclusion, we have studied, in the framework of open quantum systems, the entanglement dynamics of two uniformly accelerated two-level atoms coupled with electromagnetic vacuum fluctuations. For atoms initially in a maximally entangled state, entanglement sudden death is a general feature for accelerated atoms with a nonvanishing separation, and the decay rate of concurrence is dependent on the initial state, acceleration, interatomic separation, and polarization directions of the atoms. When both of the two atoms are initially in the excited state, we investigate the delayed sudden birth of entanglement. The maximum concurrence generated during the evolution of static atoms decreases as acceleration increases, while for accelerated atoms it may not be a monotonic function of acceleration. Both the lifetime of entanglement and the maximum concurrence generated during evolution may be affected by the atomic polarization directions. When the atoms are initially prepared in a superposition of $\ket{A}$ and $\ket{S}$, or $\ket{E}$ and $\ket{G}$, for certain initial states and interatomic separations, the existence of entanglement sudden revival and enhancement is dependent on the atomic polarizations. A comparison between the possible regions of entanglement revival and enhancement for accelerated atoms and static ones in a thermal bath shows that they do not completely overlap.

\begin{acknowledgments}
This work was supported in part by the NSFC under Grants No. 11375092, No. 11435006, and No. 11447022;  the Zhejiang Provincial Natural Science Foundation of China under Grant No. LQ15A050001; the Research Program of Ningbo University under Grants No. XYL15020 and No. xkzwl1501; and the K. C. Wong Magna Fund in Ningbo University.
\end{acknowledgments}


\begin{thebibliography}{00}


\bibitem{sch}
E. Schr\"{o}dinger, Naturwissenschaften {\bf 23}, 807 (1935).

\bibitem{information1}
M. A. Nielsen and I. L. Chuang, {\it Quantum Computation and Quantum Information} (Cambridge University Press, Cambridge, 2000).

\bibitem{information2}
D. Bouwmeester, A. Ekert and A. Zeilinger, {\it The Physics of Quantum Information } (Springer, Berlin, 2000).

\bibitem{information3}
L. Viola, S. Lloyd, and E. Knill, Phys. Rev. Lett. {\bf 83}, 4888 (1999).

\bibitem{information4}
A. Beige, D. Braun, B. Tregenna, and P. L. Knight, Phys. Rev. Lett. {\bf 85}, 1762 (2000).

\bibitem{esd1}
T. Yu and J. H. Eberly, Phys. Rev. Lett. {\bf 93}, 140404 (2004).

\bibitem{esd2}
J. H. Eberly and T. Yu, Science {\bf 316}, 555 (2007).

\bibitem{esd-3}
M. P. Almeida, F. de Melo, M. Hor-Meyll, A. Salles, S. P. Walborn, P. H. Souto Ribeiro, L. Davidovich, Science {\bf 316}, 579 (2007).

\bibitem{ent-birth}
Z. Ficek and R. Tana\'{s}, Phys. Rev. A {\bf 77}, 054301 (2008).

\bibitem{ent-bath1}
D. Braun, Phys. Rev. Lett. {\bf 89}, 277901 (2002).

\bibitem{ent-bath2}
M. S. Kim, J. Lee, D. Ahn and P. L. Knight, Phys. Rev. A {\bf 65}, 040101(R) (2002).

\bibitem{ent-bath3}
R. Tana\'{s} and Z. Ficek, Phys. Scr. {\bf T140}, 014037 (2010).

\bibitem{ent-rev}
Z. Ficek and R. Tana\'{s}, Phys. Rev. A {\bf 74}, 024304 (2006).



\bibitem{Benatti-job}
F Benatti and R Floreanini, J. Opt. B {\bf 7}, S429 (2005).

\bibitem{Unruh}
W. G. Unruh, Phys. Rev. D {\bf 14}, 870 (1976).

\bibitem{Benatti-pra}
F. Benatti and  R. Floreanini, Phys. Rev. A {\bf 70}, 012112 (2004).

\bibitem{yu-prd}
J. Zhang and H. Yu, Phys. Rev. D {\bf 75}, 104014 (2007).

\bibitem{Matsas}
A. G. S. Landulfo and G. E. A. Matsas, Phys. Rev. A {\bf 80}, 032315 (2009).

\bibitem{Doukas1}
J. Doukas and B. Carson, Phys. Rev. A {\bf 81}, 062320 (2010).

\bibitem{BLHu}
D. C. M. Ostapchuk, S.-Y. Lin, R. B. Mann and B. L. Hu, J. High Energy Phys. {\bf 07}, 072 (2012).


\bibitem{Hu-pra}
J. Hu and H. Yu, Phys. Rev. A {\bf 91}, 012327 (2015).


\bibitem{Takagi}
S. Takagi, Prog. Theor. Phys. Suppl. {\bf 88}, 1 (1986).

\bibitem{em2}
Z. Zhu, H. Yu and S. Lu, Phys. Rev. D {\bf 73}, 107501 (2006).

\bibitem{em3}
H. Yu and Z. Zhu, Phys. Rev. D {\bf 74}, 044032 (2006).

\bibitem{em4}
W. Zhou and H. Yu, Class. Quantum Grav. {\bf 29},  085003 (2012).

\bibitem{lamb1}
R. Passante, Phys. Rev. A {\bf 57}, 1590 (1998).

\bibitem{lamb2}
Z. Zhu and H. Yu, Phys. Rev. A {\bf 82}, 042108 (2010).

\bibitem{Doukas2}
J. Doukas and L. C. L. Hollenberg, Phys. Rev. A {\bf 79}, 052109 (2009).



\bibitem{Lindblad}
V. Gorini, A. Kossakowski, and E. C. G. Surdarshan, J. Math. Phys. {\bf 17}, 821 (1976).

\bibitem{Lindblad2}
G. Lindblad, Commun. Math. Phys. {\bf 48}, 119 (1976).

\bibitem{open}
H.-P. Breuer and F. Petruccione, {\it The Theory of Open Quantum Systems} (Oxford University Press, Oxford, 2002).

\bibitem{ent-states}
Z. Ficek and R. Tana\'{s}, Phys. Rep. {\bf 372}, 369 (2002).

\bibitem{concurrence1}
W. K. Wootters, Phys. Rev. Lett. {\bf 80}, 2245 (1998).

\bibitem{concurrence2}
Z. Ficek and R. Tana\'{s}, J. Opt. B: Quantum Semiclass. Opt. {\bf 6}, S90-S97 (2004).

\bibitem{BLHu12}
C. H. Fleming, N. I. Cummings, C. Anastopoulos, and B. L. Hu, J. Phys. A: Math. Theor. {\bf 45}, 065301 (2012).

\bibitem{BLHu09}
S.-Y. Lin and B. L. Hu, Phys. Rev. D {\bf 79}, 085020 (2009).


\bibitem{QFTCS}
N. D. Birrell and P. C. W. Davies, {\it Quantum Fields in Curved Space} (Cambridge University Press, Cambridge, 1982).

\bibitem{martin}
A. Pozas-Kerstjens and E. Mart\'{i}n-Mart\'{i}nez, Phys. Rev. D {\bf 94}, 064074 (2016).


\end{thebibliography}
\end{document}